\documentclass{article}
\usepackage{amssymb}

\usepackage{graphicx}
\usepackage{amsmath}

\input{tcilatex}

\begin{document}

\title{Non-extensive Random Matrix Theory - A Bridge Connecting Chaotic and Regular
Dynamics}
\author{A. Y. Abul-Magd \\
Faculty of Science, Zagazig University, Zagazig, Egypt}
\maketitle

\begin{abstract}
We consider a possible generalization of the random matrix theory, which
involves the maximization of Tsallis' $q$-parametrized entropy. We discuss
the dependence of the spacing distribution on $q$ using a non-extensive
generalization of Wigner's surmises for ensembles belonging to the
orthogonal, unitary and symplectic symmetry universal classes.

\end{abstract}

\section{\noindent Introduction}

Non-extensive statistics is by now recognized as a new paradigm for
statistical mechanical consideration \cite{tsallis}. It revolves around the
concept of Tsallis's information-entropy measure $S_{q}$, a generalization
of the Boltzmann-Gibbs-Shannon (BGS) entropy, that depends on a real
parameter $q$. This parameter, which is called entropic index, characterizes
the degree of extensivity of the system. If the system is composed of two
independent systems $A$ and $B$ such that the probability $p(A+B)=p(A)p(B),$
the entropy of the total system is given by the following pseudo-extensive
relation 
\begin{equation}
S_{q}(A+B)=S_{q}(A)+S_{q}(B)+(1-q)S_{q}(A)S_{q}(B),
\end{equation}
from which the denunciation non-extensive comes. Tsallis's entropy becomes
Shannon's measure for the particular value $q=1$. BGS entropy is the natural
entropy for modelling chaotic dynamics that suffer from exponential
sensitivity to initial condition, having at least one positive Lyapunov
exponent. Tsallis statistics provide possible way\ to describe
non-integrable systems when the sensitivity to the initial conditions
diverges less than exponentially. This has been demonstrated by studying a
great variety of systems. Among them is the logistic map \cite
{Ts1,Ts2,baldovin}, which have been used since more than two decades to
demonstrate various routes to chaos \cite{Cv}.

Quantum chaotic systems are known to be described in terms of the
random-matrix theory (RMT). It models a chaotic system by an ensemble of
random Hamiltonian matrices $H$ that belong to one of the three universal
classes, orthogonal, unitary and symplectic. The Gaussian probability
density distribution of the matrix elements $P\left( H\right) \varpropto
\exp \left[ -\eta \text{Tr}\left( H^{\dagger }H\right) \right] $ is obtained
by maximizing BGS entropy \cite{mehta,balian}. A complete discussion of the
level correlations even for these three canonical ensemble is a difficult
task. \ Most of the interesting results are obtained for the limit of $%
N\rightarrow \infty $. Analytical results have long ago been obtained for
the case of $N=2$ \cite{porter}. It yields simple analytical expressions for
the nearest-neighbor-spacing (NNS) $P(s)$, renormalized to make the mean
spacing equal one. The spacing distribution for the Gaussian orthogonal
ensemble (GOE), $P(s)=\frac{\pi }{2}se^{-\frac{\pi }{4}s^{2}}$, is known as
Wigner's surmise. It is obtained by imposing the requirement of being
invariant with respect to the base transformation \cite{porter,haake}.
Analogous expression are obtained for the Gaussian unitary and symplectic
ensembles (GUE and GSE), respectively \cite{porter,haake}, and given here
below. These expressions have been successfully applied to the NNS of
various chaotic systems. On the other hand, there are elaborate theoretical
arguments by Berry and Tabor \cite{tabor} that NNS of classically integrable
systems should have a Poissonian statistics. The Poisson distribution $\exp
(-s)$\ of the regular spectra has been proved in some cases (see, results by
Sinai \cite{sinai} and Marklof \cite{marklof}, for instance). Still its
mechanism is not completely understood. It has also been confirmed by many
numerical studies, e.g. \cite{rob}, so that the appearance of the Poisson
distribution is now admitted as a universal phenomenon in generic integrable
quantum systems. Nevertheless, it is well known that not all the regular
systems have a Poissonian NNS\ distribution. The two-dimensional harmonic
oscillator is a classical example \cite{tabor}. In this case, numerical
calculations show the NNS distribution vanishes everywhere except in a
narrow region centered at $s\approx 1$. This suggests that the spectrum has
nearly a picket-fence shape.

In this work, we suggest that the non-extensive generalization to RMT
provides a principled way to accommodate systems with mixed regular-chaotic
dynamics. We use the maximum entropy principle, with Tsallis' entropy, to
obtain the distribution functions for random-matrix ensembles belonging to
the three canonical symmetry universalities.

\section{Nonextensive RMT}

The Tsallis entropy is defined for the continuous variables entering the
Hamiltonian matrix $H$ by 
\begin{equation}
S_{q}\left[ P_{\beta }(q,H)\right] =\frac{1-\int dH\left[ P_{\beta }(q,H)%
\right] ^{q}}{q-1},
\end{equation}
where 
\begin{equation}
dH=\prod_{m\geq n}dH_{mn}^{(0)}\prod_{\gamma =1}^{\beta
-1}\prod_{m>n}dH_{mn}^{(\gamma )},
\end{equation}
while $\beta =1,2$ and 4 for the Hamiltonians having orthogonal, unitary or
symplectic symmetries. We shall refer to the corresponding ensembles as the
Tsallis orthogonal ensemble (TsOE), the Tsallis Unitary ensemble (TsUE), and
the Tsallis symplectic ensemble (TsSE). For $q\rightarrow 1$, $S_{q}\ $tends
to BGS entropy $S_{1}\left[ P_{\beta }(q,H)\right] =-\int dHP_{\beta
}(q,H)\ln P_{\beta }(q,H)$, which yields the Gaussian ensembles \cite
{mehta,balian}. There are more than one formulation of non-extensive
statistics which mainly differ in the definition of the averaging. Some of
them are discussed in \cite{wang}. We apply the most recent formulation \cite
{Ts3}. The probability distribution $P_{\beta }(q,H)$\ is obtained by
maximizing the entropy under two conditions, 
\begin{eqnarray}
\int dHP_{\beta }(q,H) &=&1, \\
\frac{\int dH\left[ P_{\beta }(q,H)\right] ^{q}\text{Tr}\left( H^{\dagger
}H\right) }{\int dH\left[ P_{\beta }(q,H)\right] ^{q}} &=&\sigma _{\beta
}^{2}
\end{eqnarray}
where $\sigma _{\beta }$ is a constant. The optimization of $S_{q}$ with
these constraints yields a power-law type for $P_{\beta }(q,H)$ 
\begin{equation}
P_{\beta }(q,H)=\widetilde{Z}_{q}^{-1}\left[ 1+(q-1)\widetilde{\eta }%
_{q}\left\{ \text{Tr}\left( H^{\dagger }H\right) -\sigma _{\beta
}^{2}\right\} \right] ^{-\frac{1}{q-1}},
\end{equation}
where $\widetilde{\eta }_{q}>0$ is related to the Lagrange multiplier $\eta $
associated with the constraint in (5) by 
\begin{equation}
\widetilde{\eta }_{q}=\eta /\int dH\left[ P_{\beta }(q,H)\right] ^{q},
\end{equation}
and\ 
\begin{equation}
\widetilde{Z}_{q}=\int dH\left[ 1+(q-1)\widetilde{\eta }_{q}\left\{ \text{Tr}%
\left( H^{\dagger }H\right) -\sigma _{\beta }^{2}\right\} \right] ^{-\frac{1%
}{q-1}}.
\end{equation}
It turns out that the distribution (6) can be written hiding the presence of 
$\sigma _{\beta }^{2}$ in a more convenient form 
\begin{equation}
P_{\beta }(q,H)=Z_{q}^{-1}\left[ 1+(q-1)\eta _{q}\text{Tr}\left( H^{\dagger
}H\right) \right] ^{-\frac{1}{q-1}},
\end{equation}
where 
\begin{equation}
\eta _{q}=\frac{\eta }{\int dH\left[ P_{\beta }(q,H)\right] ^{q}+(1-q)\eta
\sigma _{\beta }^{2}},
\end{equation}
and 
\begin{equation}
Z_{q}=\int dH\left[ 1+(q-1)\eta _{q}\text{Tr}\left( H^{\dagger }H\right) %
\right] ^{-\frac{1}{q-1}}.
\end{equation}
The non-extensive distribution (9) reduce to the statistical weight of the
Gaussian ensemble when $q=1$.

We now calculate the joint probability density for the eigenvalues of the
Hamiltonian $H$. With $H=U^{-1}XU$, where $U$\ is the global unitary group.
For this purpose, we introduce the elements of the diagonal matrix of
eigenvalues $X=$ diag$(x_{1},\cdots ,x_{N})$ of the eigenvalues and the
independent elements of $U$ as new variables. Then the volume element (3)
has the form 
\begin{equation}
dH=\left| \Delta _{N}\left( X\right) \right| ^{\beta }dXd\mu (U),
\end{equation}
where $\Delta _{N}\left( X\right) =\prod_{n>m}(x_{n}-x_{m})$ is the
Vandermonde determinant and $d\mu (U)$ the invariant Haar measure of the
unitary group \cite{mehta}. In terms of the new variables, Tr$\left(
H^{\dagger }H\right) =\sum_{i=1}^{N}x_{i}^{2}$ so that the right-hand side
of Eq. (9) is independent of the ''angular'' variables in $U$. \ Integrating
(6) over $\mu (U)$ yields the joint probability density of eigenvalues in
the form 
\begin{equation}
P_{\beta }^{(q)}(x_{1},\cdots ,x_{N})=C_{\beta }(q)\left|
\prod_{n>m}(x_{n}-x_{m})\right| ^{\beta }\left[ 1-(1-q)\eta
_{q}\sum_{i=1}^{N}x_{i}^{2}\right] ^{\frac{1}{1-q}},
\end{equation}
where $C_{\beta }$ is a normalization constant. This formula is the main
result of the proposed nonextensive generalization of RMT and, to our
knowledge, the eigenvalue distribution functions has never been expressed in
this form before. In the limit of $q=1$, it reduces to eigenvalue
distribution functions obtained in the RMT and used for further
investigation of the Gaussian ensembles.

\section{Nonextensive Wigner surmises}

Analytical expressions for the nearest-neighbor spacing (NNS) distribution
can be obtained for the $2\times 2$ matrix ensembles. For the Gaussian
ensembles, this approach leads to the well-known Wigner surmises 
\begin{equation}
p_{\beta }(s)=a_{\beta }s^{\beta }\exp \left( -b_{\beta }s^{2}\right) ,
\end{equation}
where $\left( a_{\beta },b_{\beta }\right) =\left( \frac{\pi }{2},\frac{\pi 
}{4}\right) $, $\left( \frac{32}{\pi ^{2}},\frac{4}{\pi }\right) $ and $%
\left( \frac{2^{18}}{3^{6}\pi ^{3}},\frac{64}{9\pi }\right) $ for GOE, GUE,
and GSE, respectively. They present accurate approximation to the exact
results for the case of $N\rightarrow \infty $.

We consider the non-extensive generalization of the Wigner surmises, hoping
that the results present a reasonable approximation to the physically
interesting cases of large $N$, as they do well in the case of Gaussian
ensembles. We rewrite Eq. (13) for the case of $N=2$, and introduce the new
variable\ $s=\left| x_{1}-x_{2}\right| $ and $X=(x_{1}+x_{2})/2$ to obtain 
\begin{equation}
P_{\beta }(q,s,X)=C_{\beta }(q)s^{\beta }\left[ 1+(q-1)\eta _{q}\left( \frac{%
1}{2}s^{2}+2X^{2}\right) \right] ^{-\frac{1}{q-1}}.
\end{equation}
We first consider the case of $q>1$ where no limitations are imposed on the
values of the variables $x_{1}$\ and $x_{2}$. Thus, we integrate over $X$
from $-\infty $ to $\infty $, and obtain 
\begin{equation}
P_{\beta }(q,s)=A_{\beta }(q)s^{\beta }\left[ 1+B_{\beta }\left( q\right)
s^{2}\right] ^{-\frac{1}{q-1}+\frac{1}{2}},
\end{equation}
where the quantities $A_{\beta }(q)$ and $B_{\beta }(q)$\ can easily be
expressed in terms of the factors $C_{\beta }(q)$ and $\eta _{q}$, and can
also be determined from the conditions of normalization and unit mean
spacing as 
\begin{equation}
A_{\beta }(q)=\frac{2\left[ B_{\beta }(q)\right] ^{\left( \beta +1\right)
/2}\Gamma \left( -\frac{1}{2}+\frac{1}{q-1}\right) }{\Gamma \left( \frac{%
\beta +1}{2}\right) \Gamma \left( -1-\frac{\beta }{2}+\frac{1}{q-1}\right) },
\end{equation}
and 
\begin{equation}
B_{\beta }(q)=\left[ \frac{\Gamma \left( \frac{\beta +2}{2}\right) \Gamma
\left( -\frac{3}{2}-\frac{\beta }{2}+\frac{1}{q-1}\right) }{\Gamma \left( 
\frac{\beta +1}{2}\right) \Gamma \left( -1-\frac{\beta }{2}+\frac{1}{q-1}%
\right) }\right] ^{2}.
\end{equation}
For $q=1$, Eq. (16) yields the well-known Wigner distributions $\backsim
s^{\beta }\exp (-as^{2})$, $a$ = const. As we increase $q$, the NNS
distribution takes the form of a power-law type. The quantity $B_{\beta }(q)$%
\ diverges when 
\begin{equation}
q=q_{\max }=1+\frac{2}{\beta +3}.
\end{equation}
This imposes an upper limit on allowed values $q$. It is equal to 3/2, 7/5
and 9/7 for the TsOE,\ TsUE and TsSE, respectively. The variance of this
distribution is given by 
\begin{equation}
\sigma _{\beta }^{2}=\frac{\Gamma \left( \frac{\beta +1}{2}\right) \Gamma
\left( \frac{\beta +3}{2}\right) \Gamma \left( -1-\frac{\beta }{2}+\frac{1}{%
q-1}\right) \Gamma \left( -2-\frac{\beta }{2}+\frac{1}{q-1}\right) }{\left[
\Gamma \left( \frac{\beta +2}{2}\right) \Gamma \left( -\frac{3}{2}-\frac{%
\beta }{2}+\frac{1}{q-1}\right) \right] ^{2}}.
\end{equation}
It diverges unless 
\begin{equation}
q<q_{\infty }=1+\frac{2}{\beta +4}.
\end{equation}
This imposes physical bound on the admissible values of $q$, because $\sigma
_{\beta }^{2}$ has to be finite in order to force condition (5). At values
of the entropic index exceeding $q_{\infty }$, non-extensive statistics does
not apply to the random matrix model. Figure 1 shows the evolution of the
NNS distributions as $q$ increases from 1 to $q_{\infty }$. The position of
the peak of the distribution, which is located at $s_{\beta }=\sqrt{\beta
/B_{\beta }(q)\left[ 1/(q-1)-1-\beta \right] }$, moves from $s_{1}=0.798$ to 
$s_{1}=0.368$ for TsOE, from $s_{2}=0.886$ to $s_{2}=0.408$ for TsUE and
from $s_{4}=0.940$ to $s_{4}=0.671$ for TsOE. Neither reaches 0, the peak
position of the Poisson distribution $\exp (-s)$ of the integrable systems.
The figure shows that distributions (16) evolve the shape predicted by the
corresponding Wigner surmise towards the Poisson distribution, but never
reach it. The limiting distributions are 
\begin{equation}
P_{\beta }(q_{\infty },s)=\left\{ 
\begin{array}{c}
\frac{8\pi ^{2}s}{\left( 4+\pi ^{2}s^{2}\right) ^{2}};\ \text{for TsOE} \\ 
\frac{24s^{2}}{\left( 1+4s^{2}\right) ^{5/2}};\text{ for TsUE} \\ 
\frac{1474560s^{4}}{\left( 9+64s^{2}\right) ^{7/2}};\text{ for TsSE}
\end{array}
\right.
\end{equation}
which all tend to $s^{-3}$ as $s\rightarrow \infty $. Recalling the results
obtained for the logistic map \cite{Ts2}, we may consider these
distributions as signaling the edge of chaos where the Lyapunov exponents
vanish. The Tsallis statistics cannot be extended to more regular regimes
probably because the degree of mixing then does not justify the use of the
principle of maximum entropy.

Now we consider the case of $q<1$. For this case, the distribution (9) has
to be complemented by the auxiliary condition that the quantity inside the
square bracket has to be positive. Thus, we integrate over $X$ from $-z$ to $%
z$, where $z=\sqrt{1-\frac{1}{2}(1-q)\eta _{q}s^{2}}$. We obtain 
\begin{equation}
P_{\beta }(q,s)=a_{\beta }(q)s^{\beta }\left[ 1-b_{\beta }\left( q\right)
s^{2}\right] ^{\frac{1}{1-q}+\frac{1}{2}},
\end{equation}
where 
\begin{equation}
a_{\beta }(q)=\frac{2\left[ b_{\beta }(q)\right] ^{\left( \beta +1\right)
/2}\Gamma \left( 2+\frac{\beta }{2}+\frac{1}{1-q}\right) }{\Gamma \left( 
\frac{\beta +1}{2}\right) \Gamma \left( \frac{3}{2}+\frac{1}{1-q}\right) },
\end{equation}
and 
\begin{equation}
b_{\beta }(q)=\left[ \frac{\Gamma \left( \frac{\beta +2}{2}\right) \Gamma
\left( 2+\frac{\beta }{2}+\frac{1}{1-q}\right) }{\Gamma \left( \frac{\beta +1%
}{2}\right) \Gamma \left( \frac{5}{2}+\frac{\beta }{2}+\frac{1}{1-q}\right) }%
\right] ^{2}.
\end{equation}
Figure 2 demonstrates the behavior of the NNS distributions for the three
ensembles as $q$ decreases from 1 to 0. The figure shows that each of the
three distributions evolves from that of the Wigner surmise towards a shape
describing a picket-fence spectrum. The limiting distributions are 
\begin{equation}
P_{\beta }(0,s)=\left\{ 
\begin{array}{c}
\frac{5^{3}\pi ^{2}}{2^{10}}s\left( 1-\frac{5^{3}\pi ^{2}}{2^{10}}%
s^{2}\right) ^{3/2};\ \text{for TsOE} \\ 
\frac{2^{23}}{5^{3}7^{3}\pi ^{4}}s^{2}\left( 1-\frac{2^{12}}{5^{2}7^{2}\pi
^{2}}s^{2}\right) ^{3/2};\text{ for TsUE} \\ 
\frac{2^{63}}{3^{16}5^{5}7^{5}}s^{4}\left( 1-\frac{2^{22}}{3^{6}5^{2}7^{2}}%
s^{2}\right) ^{3/2};\text{ for TsSE.}
\end{array}
\right.
\end{equation}
These distributions have their peaks at $s\cong 1$ as that of the
picket-fence distribution, but their widths are not small.

\section{Conclusion}

In summary, the NNS distribution obtained from applying Tsallis statistics
to RMT maps two routes for the transition from chaos to order. One leads
towards the integrability described by the Poisson statistics by increasing $%
q>1$, but ends at the edge of chaos. The other is followed by decreasing $q$
form 1 to 0. It leads to a picket-fence type spectrum, such as the one
obtained by Berry and Tabor \cite{tabor}\ for the two-dimensional harmonic
oscillator with non-commensurate frequencies.

In order to get a more complete picture of the non-extensive RMT, we need to
extend our analysis using ensembles of large matrices and including other
observables as level-number variance $\Sigma ^{2}$ and spectral rigidity $%
\Delta _{3}$. Work on these lines is in progress.

\begin{description}
\item  {\LARGE Figure Caption}

\item  Fig.1. The evolution of the NNS distribution for the three symmetry
universalities from the Wigner form towards the Poissonian (both shown by
solid curves) by varying the entropic index in the range of $1\leq q\leq
q_{\infty }$. The dashed curves are for distributions having second moments
of 2, equal to the Poissonian. The dotted curves are for the values $%
q=q_{\infty }$ explicitly given by Eqs. (22).\pagebreak 

\item  Fig.2. The evolution of the NNS distribution for the three symmetry
universalities from the Wigner form (solid curves) towards the picket-fence
shape by varying the entropic index in the range of $0\leq q\leq 1$. The
dashed curves are for distributions having $q=0.5$. The dotted curves are
for the values $q=0$, where the distributions are explicitly given by Eqs.
(26).
\end{description}

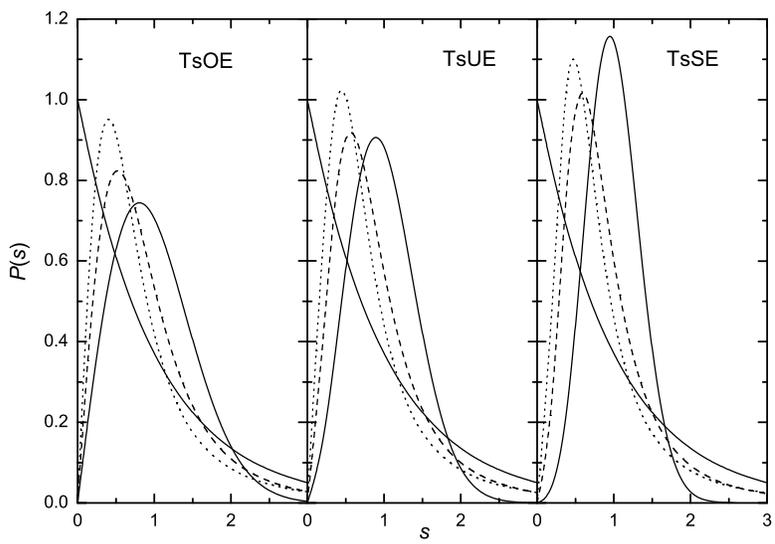
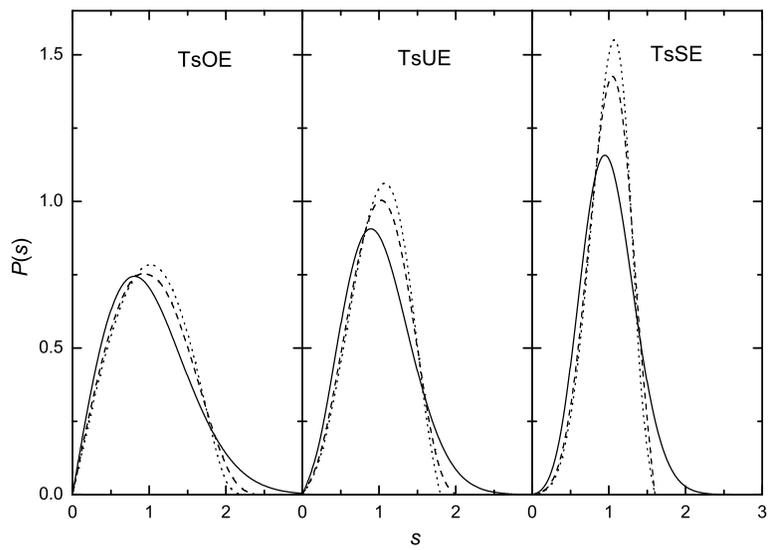
\end{document}